\documentclass[prd, twocolumn, nofootinbib,showkeys,showpacs, floatfix]{revtex4-1}

\usepackage[mathscr]{euscript}
\usepackage{amsmath}

\usepackage{graphicx}
\usepackage{dcolumn}
\usepackage{bm}
\usepackage{epsfig}
\usepackage{amssymb,latexsym,mathrsfs}
\usepackage{graphicx}
\usepackage{color}
\usepackage{hyperref}
\usepackage{float}
\usepackage{diagbox}
\usepackage{physics}
\usepackage{braket}
\usepackage{tensor}

\definecolor{vividviolet}{rgb}{0.62, 0.0, 1.0}
\definecolor{amaranth}{rgb}{0.9, 0.17, 0.31}
\definecolor{palatinateblue}{rgb}{0.15, 0.23, 0.89}
\definecolor{brightpink}{rgb}{1.0, 0.0, 0.5}
\definecolor{cornflowerblue}{rgb}{0.39, 0.58, 0.93}
\definecolor{deepcarminepink}{rgb}{0.94, 0.19, 0.22}
\definecolor{radicalred}{rgb}{1.0, 0.21, 0.37}

\hypersetup{ linktoc=all,
    colorlinks, linkcolor={palatinateblue},
    citecolor={brightpink}, urlcolor={amaranth}
}

\newcommand{\be}{\begin{equation}}
\newcommand{\ee}{\end{equation}}
\newcommand{\bs}{\begin{split}} 
\newcommand{\bea}{\begin{eqnarray}}
\newcommand{\eea}{\end{eqnarray}}

\newcommand{\al}{\alpha} 
 
\newcommand{\kp}{\kappa}
\newcommand{\eps}{\epsilon} 
\newcommand{\tst}{t_\star}

\renewcommand{\d}[1]{\ensuremath{\operatorname{d}\!{#1}}}

\newcommand{\bo}{\raise-1mm\hbox{\Large$\Box$}}

\newcommand{\bes}{\begin{subequations}}
\newcommand{\ees}{\end{subequations}}

\begin{document} 
\title{Quantum power: a Lorentz invariant approach to Hawking radiation}
\author{Michael R.R. Good${}^{1,2}$}
\email{michael.good@nu.edu.kz}
\author{Eric V. Linder${}^{2,3}$}
\email{evlinder@lbl.gov}
\affiliation{${}^1$Physics Department, Nazarbayev University,
Nur-Sultan, Kazakhstan\\
${}^2$Energetic Cosmos Lab, Nazarbayev University, 
Nur-Sultan, Kazakhstan\\
${}^3$Berkeley Center for Cosmological Physics \& Berkeley Lab, University of California, Berkeley, CA, USA 
}

\begin{abstract} 
Particle radiation from black holes has an observed emission power depending on the surface gravity $\kappa = c^4/(4GM)$ as    
\be\nonumber
P_{\textrm{black hole}} \sim \frac{\hbar \kappa^2}{6\pi c^2} = \frac{\hbar c^6}{96\pi G^2 M^2}\,,\ee
while both the radiation from accelerating particles and moving mirrors (accelerating boundaries) obey similar relativistic Larmor powers,
\be\nonumber
P_{\textrm{electron}}= \frac{q^2\alpha^2}{6\pi \epsilon_0 c^3}\,, \quad P_{\textrm{mirror}}  =\frac{\hbar \alpha^2}{6\pi c^2}\,, 
\ee 
where $\al$ is the Lorentz invariant proper acceleration. 
This equivalence between the Lorentz invariant powers suggests 
a close relation that could be used to understand black hole radiation. 
We show that an accelerating mirror with a prolonged metastable 
acceleration plateau can 
provide a unitary, thermal, energy-conserved analog model for black hole decay.  
\end{abstract}

\maketitle


\section{Introduction} 

The Equivalence Principle teaches us that gravitation, acceleration, and curvature are equivalent. Moreover we know that external effects on quantum fields creates particles, and this ties together black hole particle production, thermal baths observed by accelerating observers, and moving mirror acceleration radiation, e.g.\ the Hawking \cite{Hawking:1974sw}, Unruh \cite{unruh1976}, and Davies-Fulling \cite{Davies:1976hi} effects. However, we also know that constant acceleration is insufficient: an electron sitting on a laboratory table in an eternal constant gravitational field of the Earth will not radiate. In the same way, an eternally exactly uniformly accelerating accelerated boundary (moving mirror) will not emit energy to an observer at infinity, e.g. \cite{Ford:1982ct}. 

Another aspect of great interest \cite{Chen:2015bcg} is that  asymptotically static mirrors preserve unitarity and information \cite{Chen:2017lum}. We explore a model that merges these two regimes of uniform acceleration and zero acceleration and show that this system can radiate particles for an extended time with constant power. The system will not only preserve information but emit thermal energy, conserve total radiated energy, and emit finite total particles, without infrared divergence. This model can serve as an analog for complete black hole evaporation.

Related explorations are not without precedent.  Black hole evaporation has close acceleration analogs \cite{Davies:1974th} including moving mirror models \cite{DeWitt:1975ys,Davies:1977yv}.  
Asymptotic infinite acceleration trajectories \cite{carlitz1987reflections} can evolve to eternal thermal equilibrium solutions \cite{Reyes}; asymptotic 
constant velocity (zero acceleration) 
can give information preserving quasi-thermal solutions describing black hole remnant models (e.g.\ \cite{Chen:2014jwq,Good:2016atu}). 
Unitary complete black hole evaporation models are characterized by asymptotic zero-velocity mirrors (e.g.\ \cite{Walker_1982, Good:2019tnf}). 

Entanglement entropy \cite{Holzhey:1994we}, and hence information, is tied directly to the mirror trajectory \cite{Akal:2020twv}.  However, 
the distant observer detects the radiated power, not the entropy. 
We investigate the connection between these   
for complete black hole evaporation  
via the analog case of uniform acceleration.  

Uniform acceleration mirrors are generally thought to emit zero energy \cite{Birrell:1982ix,Good:2021iny}.  In our case, we will explore metastable uniform acceleration, where 
there is an extended but finite period of constant power emission. 
We will confirm that the stress is zero during this plateau period but find that the power is not.  
The model presented here will preserve information, evolve to thermal equilibrium, and conserve emitted energy, providing an analog for a black hole that completely evaporates away into radiation. 

In Sec.~\ref{sec:acc} we exhibit the mirror dynamics of acceleration and velocity with the desired properties, 
leading in Sec.~\ref{sec:entropy} 
to evolution with quantum purity (information preservation) from the finite entanglement entropy with a Page turnover. Section~\ref{sec:stress} computes the quantum Larmor power and total energy radiated, linking the mirror parameters with black hole properties. 
We conclude in Sec.~\ref{sec:concl},  highlighting the unitarity and thermality of the analog models for black hole evaporation.

\section{Acceleration \& Velocity} \label{sec:acc} 

We seek a mirror acceleration that dies to zero at $\pm\infty$ (to preserve information) 
and has a constant plateau at some maximum acceleration (for metastable thermal power). We can arrange the maximum to be at time $t=0$, for example. We would also like to be able to adjust the duration of the plateau, to study the scaling. A simple model is 
\be 
\al=\al_0\, e^{-(t/\tst)^j}\,{\rm sgn}(t)\,. \label{eq:altrs} 
\ee 
The metastable plateau runs over  
$|t|\lesssim\tst$; at $t=\tst$ the acceleration falls to $1/e$ 
of its maximum value $\al_0$. As a foreshadowing, we expect the 
power emission to determine the black hole lifetime, 
$\dot M\sim P\sim M^{-2}$, where $M$ is the black hole mass, 
so we anticipate a successful analog model will have 
$\tst\sim M^3$. 

We take $j$ to be a positive even integer so that $\al$ will die to zero for $t\to\pm\infty$. 
Large $j$ gives a flatter plateau and a steeper fall off to approach zero. 
For example, the acceleration plateau stays within a fraction $\eps$ of the maximum for $|t|<\tst \eps^{1/j}$ so for $j=4$ (8) it is within 1\% of maximum out to $|t|<0.32\tst$ ($0.56\tst$). 
The limit $j\to\infty$ gives a box function for the plateau. 
This approaches equilibrium emission on the plateau. The sign 
flip (change in direction) in acceleration at $t=0$ is so the mirror comes back to 
rest (not merely inertial, but static) at future infinity. Since 
power depends on $\al^2$, the sign flip does not affect the 
power detected by a distant observer. 
(One can easily regularize the sign flip through use 
of a tanh transition without affecting the results.) 

\begin{figure}[h]
    \centering
    \includegraphics[width=3.0in]{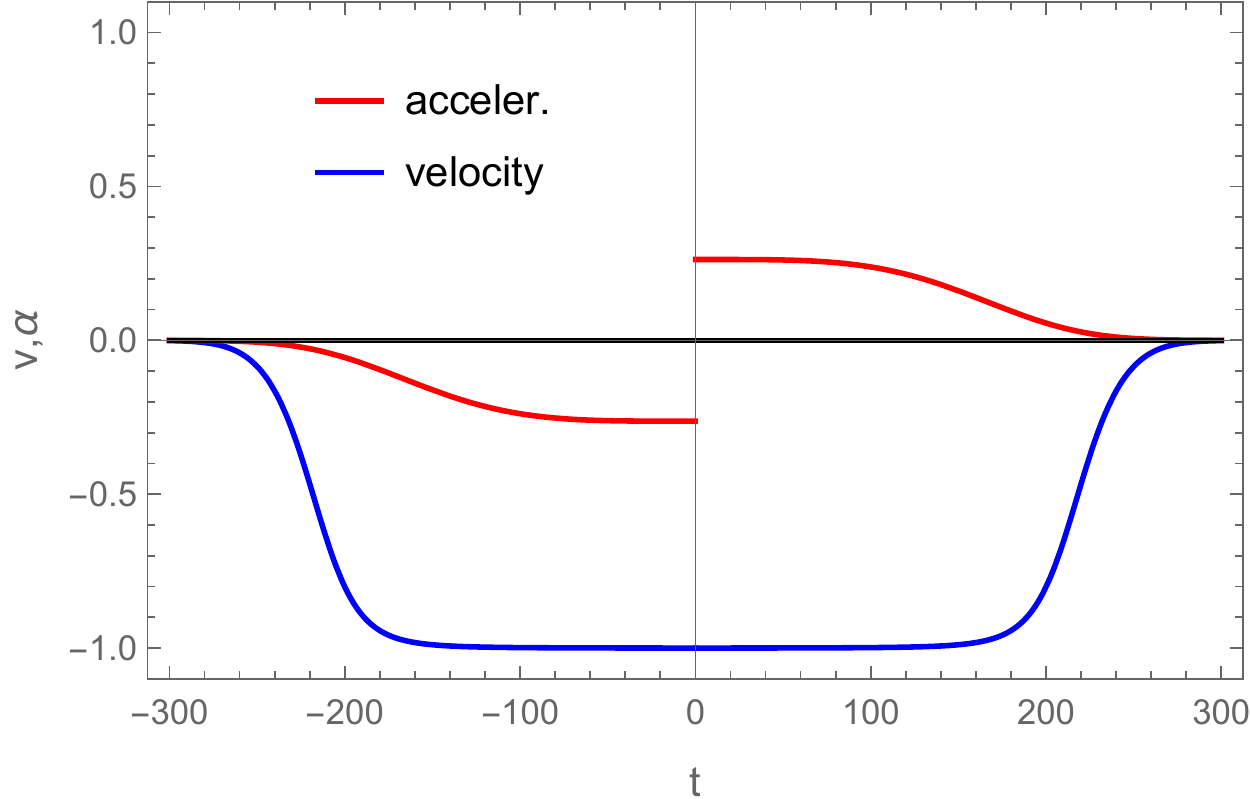}
    \caption{
The proper acceleration (red curve) of the mirror starts to the left (negative by convention), reaches a maximum magnitude ($1/4M$ as $j\to \infty$), and has a sign (direction) change at $t=0$. 
The velocity (blue curve) of the (1+1)D mirror trajectory is always to the left (by convention); the mirror starts from zero speed, approaches the speed of light, then 
finally comes to rest. Here we plot for $M = 1$ and $j=4$.  }
    \label{Fig1}
\end{figure}

The mirror velocity $v$ comes from the acceleration via the rapidity $\eta$, by $v=\tanh\eta$ and 
\be 
\sinh\eta(t)\equiv\int_{-\infty}^t dt'\,\al(t')=\al_0\tst \frac{\Gamma(1/j,(t/\tst)^j)}{j}\,, \label{eq:sinheta}
\ee 
where $\Gamma$ is the incomplete Gamma function. The velocity 
smoothly goes from 0 to a maximum near the speed of light and 
back to 0, without changing sign. 
The maximum velocity will be reached at $t=0$, where the 
incomplete Gamma function becomes a complete one, so 
\bea 
\sinh\eta(t=0)&=&\al_0 \tst\frac{\Gamma(1/j)}{j}\equiv Q \label{eq:q}\\ 
v_{\rm max}&=&\left[1+Q^{-2}\right]^{-1/2}\,.
\eea 
When $j\to\infty$, then $Q=\al_0\tst$. 
Note the maximum Lorentz boost factor $\gamma_{\rm max}=(1+Q^2)^{1/2}$. 
Figure~\ref{Fig1} shows the acceleration and the resulting 
velocity.

\section{Entropy \& Unitarity} \label{sec:entropy} 

\begin{figure}[h]
    \centering
    \includegraphics[width=3.0in]{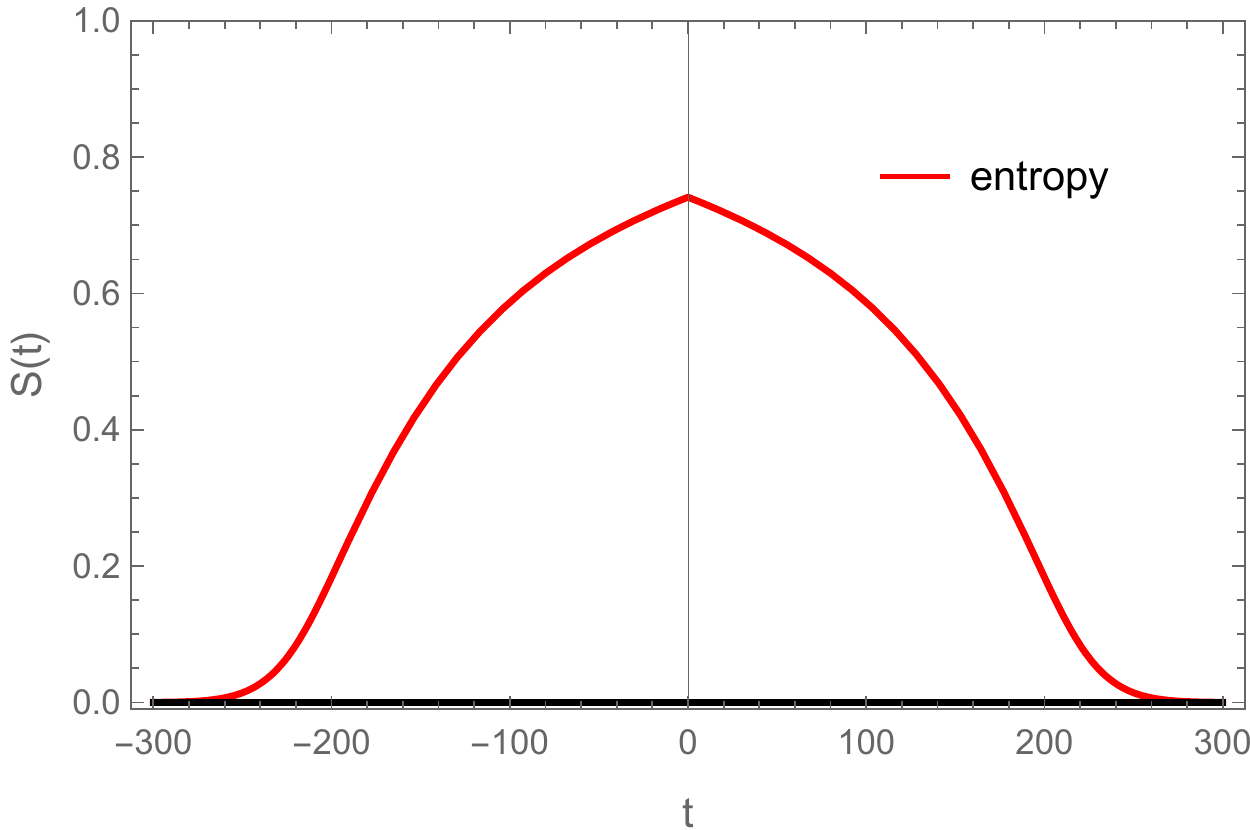}
    \caption{Page curve turn-over of the von Neumann geometric entanglement entropy, Eq.~(\ref{entropy}), with asymptotic zeros and no divergence.  This scalar measure of information demonstrates the model has no information loss by construction. Here $M=1$ and $j=4$ as in Figure \ref{Fig1}. 
}
    \label{Fig2}
\end{figure}

Before proceeding further, let us establish this is 
a unitary analog model by observing that the entanglement entropy does not diverge, as expected for a mirror with asymptotic 
static end states \cite{Reyes}. From Eq.~\eqref{eq:sinheta},  
\be 
S(t) \equiv \frac{\eta(t)}{6} = \frac{1}{6} \sinh^{-1}  
\left( \al_0\tst \frac{\Gamma(1/j,(t/\tst)^j)}{j}\right)\,.\label{entropy} 
\ee 
The entropy is asymptotically zero (no divergence), which signals purity.  That is, in the limit $t\rightarrow \pm\infty$, $S\to 0$.  This ensures that every field mode reflects to the observer.  Without loss of field modes past a horizon, the model preserves quantum information during time evolution of the vacuum state \cite{wilczek1993quantum}. 
Figure~\ref{Fig2} exhibits the expected Page curve 
turn-over.

We can push this further, deriving thermodynamic entropy from entanglement entropy in the analog context. 
To reach the thermodynamic regime, we apply two equilibrium limits: flatness $j\to\infty$, and central time $t \to 0^-$.  The value of the rapidity here is given by $\sinh\eta_\star = \alpha_0 t_* = 12\pi M^2 = 3\pi/(4\kp^2)$  
(as seen for $t_*$ in Sec.~\ref{sec:stress} in the $j\to\infty$ limit), with $\kp$ the surface gravity.  The entanglement is then
\be 
S=\frac{2\times1}{6}\sinh^{-1}(\al_0\tst)=\frac{1}{3}\sinh^{-1}\left(\frac{3A}{4}\right)\,.
\ee 
In the first step, we have translated from dynamics to entanglement using the rapidity-entropy relation \cite{Good_2015BirthCry} and accounted for (3+1) dimensions \cite{Zhakenuly:2021pfm}, where the additive modulus entanglement entropy is twice the one-sided entropy of a mirror in (1+1) dimensions, $S = 2(\eta/6)$.  In the second step, we have written the (3+1) dimensional entanglement entropy in terms of the area of the analog black hole, $A=\pi/\kappa^2$, which illustrates thermodynamic entropy, $S=A/4$, of the gravitational analog in the geometric limit $A \to 0$.

\section{Power \& Total Energy} \label{sec:stress}

The relativistic Larmor form  for power, familiar from electrodynamics \cite{Jackson:490457}, 
also applies to the energy radiated from accelerating mirrors \cite{Zhakenuly:2021pfm}. In the latter case 
\be 
P(t) = \frac{\hbar \alpha^2(t)}{6\pi c^2}\,, 
\ee
where $\alpha$ is the frame-invariant proper acceleration.  This measure is a good candidate for what the observer detects at asymptotic infinity.  It is a Lorentz invariant corresponding to the emitted radiation from both sides of a (1+1)D moving mirror,  
as well as the emitted power for a (3+1)D moving mirror.  

Thus the power for the corresponding 
(3+1) dimensional situation of Eq.~\eqref{eq:altrs} is 
\be 
P(t) = \frac{\hbar\al_0^2}{6\pi c^2}\ e^{-2(t/\tst)^j} =\frac{c^5}{G}\,\frac{1}{6\pi}\left(\frac{\al}{\alpha_{\rm Pl}}\right)^2\,, \label{power} 
\ee 
where here we explicitly show the ``Planck power'' $c^5/G$ and 
``Planck acceleration'' $\alpha_{\rm Pl}\equiv c/t_{\rm Pl}$, 
although 
usually elsewhere we work in units where $\hbar=c=G=1$. 

\begin{figure}[h]
    \centering
    \includegraphics[width=3.0in]{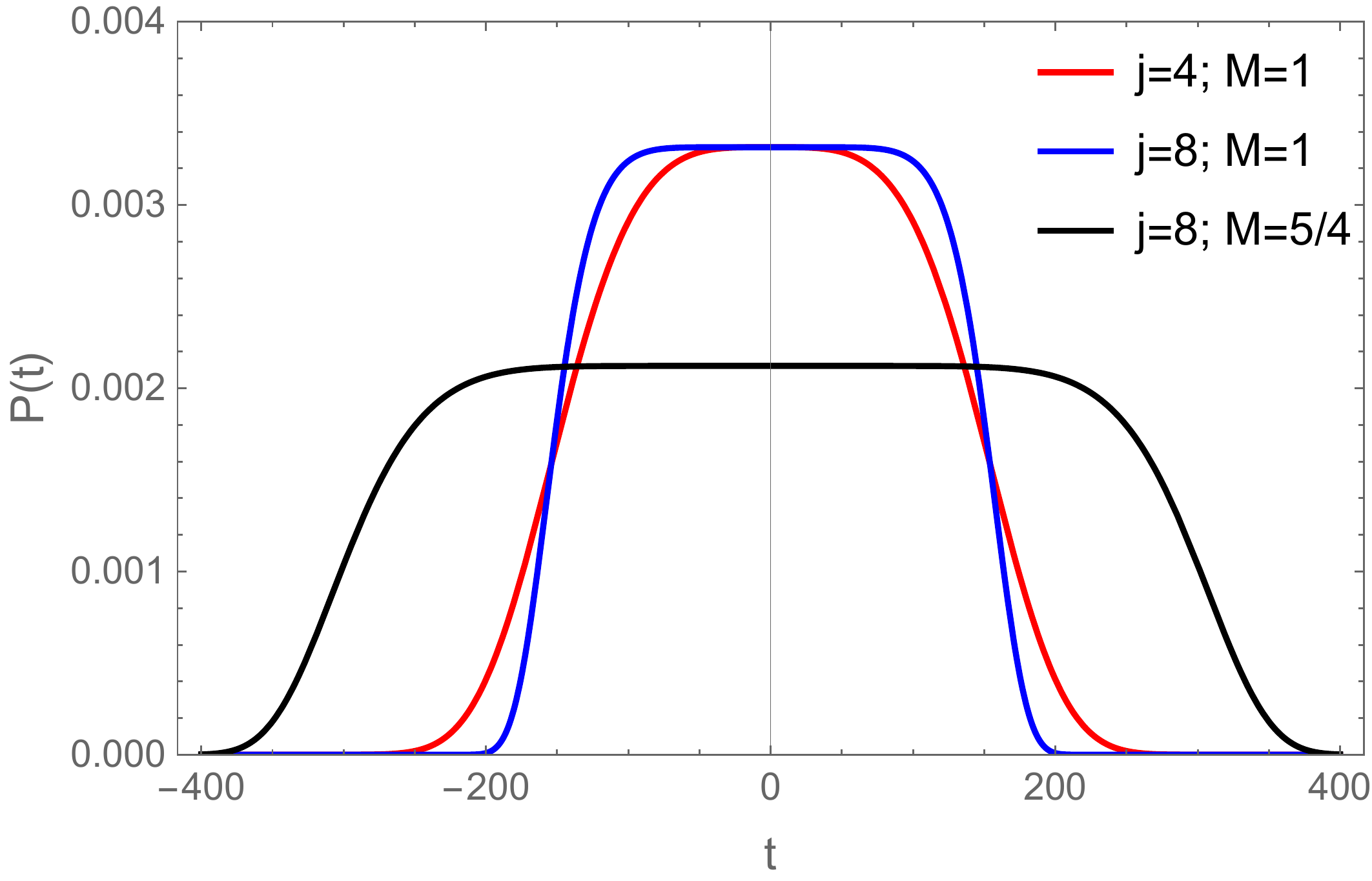}
    \caption{The acceleration plateau results in a (3+1)D power emission exhibiting an equilibrium emission plateau, with asymptotic finality (as $t\to\pm\infty$) corresponding to complete evaporation. 
}
    \label{Fig3}
\end{figure}

Figure~\ref{Fig3} shows the emitted power $P(t)$  
with its plateau, increasingly in equilibrium 
for large values of $j$, and vanishing at asymptotically 
early and late times. 
The asymptotically zero emission signals the end of evaporation, e.g.\ of the (analog) black hole, and a resulting finite total energy.

To find the total energy emitted by the evaporated (3+1)D black hole analog, one can integrate the power over coordinate time, $E = \int_{-\infty}^{+\infty} P(t) \, dt$.  The result is 
\be 
E=\frac{\al_0^2\tst}{3\pi}\frac{\Gamma(1/j)}{j\,2^{1/j}}\,,  \label{eq:energy} 
\ee 
where now we have a complete Gamma function. 
As $j\to\infty$, the second fraction goes to 1. 

We can use this to fix $\tst$ in terms of $E$ and $\al_0$. 
In particular, in the black hole context it is natural to 
take the correspondence that 
the total energy emitted is the mass $M$ of the black hole, 
and the acceleration during the equilibrium emission is 
the surface gravity, i.e.\ $\al_0=\kappa\equiv 1/(4M)$. 
This then implies that 
\bea 
\tst&=&\frac{3\pi E}{\al_0^2}\frac{j\,2^{1/j}}{\Gamma(1/j)}\\ 
&\to&48\pi M^3 \frac{j\,2^{1/j}}{\Gamma(1/j)} \to 48\pi M^3=\frac{3\pi}{4\kappa^3}\,, \label{eq:tst} 
\eea 
The first right arrow takes the correspondence, and the second 
arrow takes the $j\to\infty$ limit. We see that as expected 
from the correspondence with black hole decay, the characteristic 
time scale $\tst\sim M^3$. (In the mirror model, the radiation 
does not stay thermal beyond $\tst$ so the total decay time is 
not defined for $j$ finite.) Note that under this correspondence, 
$Q=12\pi M^2$ in Eq.~\eqref{eq:q}. 
We used this expression in 
Sec.~\ref{sec:entropy} to evaluate 
the thermodynamic entropy. 

Conversely, we can write $\al_0^2=3\pi E/\tst$ in the limit 
and find 
\be 
P=\frac{c^5}{G}\frac{GM}{2(c\tst)c^2}\,, 
\ee 
i.e.\ there is a fundamental limit in that the decay time 
$\tst$ must be long enough that the Planck power is not 
exceeded: one cannot radiate away the energy in shorter 
than a light crossing time.

This accelerating mirror model therefore provides an analog 
to the concept that a physical black hole emits finite energy,  that this energy is the total mass of the black hole itself for complete evaporation, the evaporation time $\sim M^3$, and 
the entropy $\sim A/4$, consistent with surface gravity $\sim 1/(4M)$ in the thermodynamic limit.

\section{Conclusion}\label{sec:concl} 

We present an analog model using the dynamical Casimir 
effect for accelerating boundaries (mirrors) to describe 
black hole complete evaporation. The approach uses the 
quantum relativistic Larmor formula for Lorentz invariant radiative power, $P= \hbar\alpha^2/6\pi c^2$ in terms of the  
proper acceleration $\alpha$. The Larmor formula works 
equivalently for (mirror) acceleration and (black hole) 
surface gravity. 

The accelerating boundary correspondence exhibits the 
desirable characteristics of unitarity, thermality 
(equilibrium emission), and energy conservation. The finite 
entropy with Page turnover preserves information. The model 
can thus describe a black hole that completely evaporates away in a physically reasonable manner. 

From a static state, the mirror accelerates to a velocity 
that can approach the speed of light (and the maximum 
rapidity is closely related to entropy, and black hole 
mass), before asymptotically becoming static again. 
While the specific model is of a mirror instantaneously 
reversing acceleration direction, a simple (e.g.\ tanh) 
regularization works in the same way. 

The metastable plateau becomes flatter, more in equilibrium, 
as the superGaussian parameter $j$ increases. While the 
formal limit of equilibrium is $j\to\infty$, even for $j=4$ 
the plateau is flat to 1\% for an extended period. In the 
limit there is a clear correspondence between the mirror  
acceleration and black hole surface gravity (and hence mass), 
total energy radiated and black hole total energy (mass), 
entropy and black hole area, and Larmor power and Hawking 
power. 

While for a unitary model radiation episodes must occur with both 
positive and negative energy fluxes (and zero flux in the  
exact constant acceleration limit), the power always 
remains positive. As a Lorentz 
invariant, power is an interesting and mostly unstudied 
avenue by which to approach the correspondence with black 
holes. 

Future work could explore the approach to equilibrium 
and the development of the particle spectrum.  
Conventional computation of the beta Bogolyubov 
coefficients is not tractable here, but the asymptotic 
static states guarantee there will be no infrared divergence 
(no black hole remnant) and a finite total number of 
particles emitted. Alternative attempts at the spectrum 
may yield insights into the particle production and other physics of black hole radiation.

\acknowledgments 

Funding from the FY2021-SGP-1-STMM Faculty Development Competitive Research Grant No. 021220FD3951 at Nazarbayev University is acknowledged. This work is supported in part by the Energetic Cosmos Laboratory. EL is supported in part by the U.S.\ Department of Energy, Office of Science, Office of High Energy Physics, under contract no.\ DE-AC02-05CH11231.

\bibliography{main}   

\end{document}